\documentclass[aps,prb,twocolumn,groupedaddress,showpacs,amsmath]{revtex4}

\usepackage{graphicx,amssymb}

\begin{document}

%%%%%%%%%%%%%%%%%%%%%%%%%%%%%%%%%%%%%%%%%%%%%%%%
%
% FRONTMATTER %

\title{$E$-band excitations in the magnetic Keplerate molecule Fe$_{30}$}

\author{O. Waldmann}
 \email[E-mail: ]{waldmann@iac.unibe.ch}
 \affiliation{Department of Chemistry and Biochemistry, University of Bern, CH-3012 Bern, Switzerland}

\date{\today}

\begin{abstract}
The low-temperature excitations in the magnetic Keplerate molecule Fe$_{30}$ as calculated by linear
spin-wave theory (SWT), modified linear SWT, and spin-level mean-field theory (SLMFT), are compared to the
recent inelastic neutron scattering results by Garlea et al. [Phys. Rev. B {\bf 73}, 024414 (2006)]. SLMFT
reproduces a part of the experimental spectrum rather well, but not all of it. SWTs yield a small fraction of
the $E$-band excitations and hence are not capable of a complete description of the excitation spectrum.
\end{abstract}

%33.15.Kr,   % magnetic moments & susceptibility of molecules
%71.70.-d,   % Level splitting and Interactions
%71.70.Gm,   % Exchange Interactions
%75.10.Jm,   % Quantized spin models
%75.30.Et,   % Exchange and Superexchange interactions
%75.50.Ee,   % Antiferromagnetics
%\keywords{}

\maketitle

%%%%%%%%%%%%%%%%%%%%%%%%%%%%%%%%%%%%%%%%%%%%%%%%
%
% INTRODUCTION %

The subject of the elementary spin excitations in finite antiferromagnetic (AFM) Heisenberg spin clusters has
become of much interest recently, as for many experimentally available molecular nanomagnets, such as the AFM
wheels or the Keplerate molecule Fe$_{30}$, the Heisenberg spin Hamiltonian
\begin{equation}
\label{eqn:H}
 \hat{H} = -\sum_{i,j} J_{ij}{ \hat{\textbf{S}}_i \cdot \hat{\textbf{S}}_j }
\end{equation}
is the appropriate starting point for a discussion.\cite{Muller01,OW_CCR} $J_{ij}$ measures the exchange
interaction between spins $i$ and $j$, and $S_i$ is the length of spin $i$. The number of spin centers in the
cluster will be denoted as $N$.

For a number of AFM Heisenberg clusters, the low-lying energy spectrum exhibits a rather particular
structure: As function of total spin $S$, the lowest-lying levels form a set of rotational bands (RBs) for
which $E(S) \propto S(S+1)$.\cite{Anderson52,Bernu92,Schnack00,OW_SPINDYN} The lowest set of bands is denoted
as $L$ band; the higher-lying ones as $E$ band (and the remaining states as quasi
continuum).\cite{OW_SPINDYN} The $L$ band is related to a quantized rotation of the N\'eel vector. For
bipartite systems, the $E$ band corresponds to (discrete) AFM spin-wave excitations. This picture of the
excitations has been confirmed experimentally for a number of bipartite systems;\cite{OW_CCR} in particular
the AFM wheels.\cite{OW_Cr8}

Also for the molecule Fe$_{30}$ a RB structure was conjectured.\cite{Schnack01} This system consists of an
icosidodecahedral arrangement of 30 spin-5/2 centers and is characterized by three AFM
sublattices.\cite{Muller99} For the existence of the $L$ band in this molecule solid evidence from both
theory and experiment is available.\cite{Muller01,Exler03} Concerning the $E$ band, however, the situation is
less clear. This brief report is motivated by two recent studies, a study of the excitations using modified
spin-wave theory,\cite{Cepas05} and an inelastic neutron scattering (INS) experiment, where the observed
scattering intensity was related to the $E$-band excitations.\cite{Garlea06} In the following, the spin
excitations in Fe$_{30}$ as calculated by linear spin-wave theory (LSWT), modified linear spin-wave theory
(mLSWT), and spin level mean-field theory (SLMFT), as well as observed experimentally by INS are compared.

%%%%%%%%%%%%%%%%%%%%%%%%%%%%%%%%%%%%%%%%%%%%%%%%
%
% Methods %

For the SWT calculations, Ref.~[\onlinecite{Cepas05}] was closely followed, albeit with a different numerical
implementation. The classical ground state of $\hat{H}$ (Ref.~\onlinecite{Luban01}) is determined by the
conditions $\textbf{S}_i \times \sum_{j} J_{ij} \textbf{S}_j  = 0$ for each $i$,\cite{Schmidt03} which are
solved numerically by iterating $\textbf{S}_i = S_i \sum_{j} J_{ij} \textbf{S}_j / |\sum_{j} J_{ij}
\textbf{S}_j|$ until convergence.\cite{HJthanks} Local coordinate frames are introduced at each spin site
such that the local $z$-axes coincide with the directions of the classical spins. Writing
$\tilde{\textbf{S}}_i = (\hat{S}_i^+,\hat{S}_i^-,\hat{S}_i^z)^T$ for the spherical spin operators in the
local frames, the Hamiltonian becomes
\begin{equation}
 \hat{H}= -\sum_{ij} J_{ij}
 \tilde{\textbf{S}}_i \cdot \textbf{U}_{ij}
\cdot \tilde{\textbf{S}}_j.
\end{equation}
Expressions for the elements $U_{ij}^{\nu\mu}$ ($\nu,\mu = +,-,z$) are given in [\onlinecite{Walker63}]. The
spin operators (in the new frames) are expressed by Holstein-Primakoff bosons, $\hat{S}_i^+ \approx \sqrt{2
S_i} a_i$, $\hat{S}_i^z = S_i - a_i^\dag a_i$, yielding $\hat{H}_B = -\sum_{ij} J_{ij} \hat{C}^{(2)}_{ij}$
with
\begin{eqnarray}
 \label{eqn:C_SUS}
 \hat{C}^{(2)}_{ij} &=& (S_i S_j - S_j a_i^\dag a_i - S_i a_j^\dag a_j ) U_{ij}^{zz}
  + 2 \sqrt{S_i S_j}
  \cr && \times
  ( U_{ij}^{++} a_i a_j + U_{ij}^{--} a_i^\dag a_j^\dag
  + U_{ij}^{+-} a_i a_j^\dag + U_{ij}^{-+} a_i^\dag a_j ).\nonumber
\end{eqnarray}
This real-space bosonic Hamiltonian is Bogliubov diagonalized numerically. We follow
Ref.~[\onlinecite{Colpa78}] and introduce $\textbf{a}^\dag =
(a_1^\dag,a_2^\dag,\ldots,a_N^\dag,a_1,a_2,\ldots,a_N)$, yielding
\begin{equation}
\label{eqn:HB2}
  \hat{H}_B = E_0 + \textbf{a}^\dag \cdot \textbf{D} \cdot \textbf{a},
\end{equation}
with the $2N$$\times$$2N$ matrix
\begin{equation}
\label{eqn:D}
 \textbf{D} = \left(\begin{array}{cc} \textbf{A} & \textbf{B} \\ \textbf{B}^* & \textbf{A}^* \end{array} \right)
\end{equation}
and the $N$$\times$$N$ matrices $\textbf{A}$ and $\textbf{B}$, which depend on $S_i$, $J_{ij}$ and
$U_{ij}^{\nu\mu}$. The properties $\textbf{A}^\dag = \textbf{A}$ and $\textbf{B}^T = \textbf{B}$ ensure the
hermiticy of $\textbf{D}$, and $\hat{H}_B$. The Bogliubov diagonalization (or, in the language of
[\onlinecite{Colpa78}], para-diagonalization) introduces $N$ new bosons $c_k$ via $\textbf{c}^\dag =
\textbf{a}^\dag \textbf{V}^\dag$. The $2N$$\times$$2N$ matrix $\textbf{V}$ obeys $\textbf{V}^\dag
\check{\textbf{1}} \textbf{V} = \check{\textbf{1}}$ with $\check{\textbf{1}}$ =
diag$(\textbf{1},\textbf{-1})$ ($\textbf{1}$ is the $N$$\times$$N$ unit matrix). This ensures the boson
character of the $c_k$. Para-diagonalization corresponds to finding a $\textbf{V}$ with
$(\textbf{V}^\dag)^{-1}\cdot \textbf{D} \cdot \textbf{V}^{-1} = {1\over 2} \textbf{E}$, where $\textbf{E}$ =
diag$(E_1,E_2,\ldots,E_N,E_1,E_2,\ldots,E_N)$. The para-diagonalized Hamiltonian then reads
\begin{equation}
\label{eqn:HB2diag}
  \hat{H}_B = E_0 + \sum_k E_k \left( c_k^\dag c_k + \frac{1}{2}\right).
\end{equation}
Numerical algorithms for determining $\textbf{E}$ and $\textbf{V}$ are given in [\onlinecite{Colpa78}]. The
procedure so far corresponds to linear SWT (LSWT), i.e., $\hat{H}_B^{(LSWT)} \equiv \hat{H}_B$. The
ground-state and excitation energies are given by $E^{(LSWT)}_0  = E_0 + \frac{1}{2} \sum_k E_k$ and
$E^{(LSWT)}_k = E_k$, respectively.

Standard SWTs, such as LSWT, start from the assumption of an ordered, symmetry-broken ground state, which is
obviously incorrect for finite clusters. In the modified SWTs,\cite{Taka87,Hirsch89,Cepas05} spin-rotational
invariance is restored "by hand" by enforcing zero on-site magnetizations, $\langle \hat{S}^z_i\rangle = S_i
- \langle a_i^\dag a_i\rangle = 0$ ($\langle.\rangle$ denotes the quantum expectation value in the ground
state). $N$ Lagrange multipliers $\mu_i$ are introduced, and the boson Hamiltonian of modified linear SWT
(mLSWT) becomes
\begin{equation}
\label{eqn:HBm}
  \hat{H}_B^{(mLSWT)} = \hat{H}_B + \sum_i \mu_i (a_i^\dag a_i - S_i).
\end{equation}
With $A'_{ij} = A_{ij} + (\mu_i/2) \delta_{ij}$ and $E_0' = E_0 - \sum_i \mu_i (S_i + \frac{1}{2})$, also
$\hat{H}_B^{(mLSWT)}$ assumes the form of Eq.~(\ref{eqn:HB2}). Para-diagonalization yields
$\hat{H}_B^{(mLSWT)} = E'_0 + \sum_k E'_k ( c_k'^\dag c'_k + \frac{1}{2})$. The ground-state and excitation
energies, $E^{(mLSWT)}_0 = E'_0 + \frac{1}{2} \sum_k E'_k$ and $E^{(mLSWT)}_k = E'_k$, implicitly depend on
the Langrange multipliers, which are found numerically by solving $\hat{H}_B^{(mLSWT)}$ iteratively until
$\langle a_i^\dag a_i\rangle = S_i$ for each $i$.

A further approach, called spin-level mean-field theory (SLMFT), was explored. The RB structure assumed for
Fe$_{30}$ implies approximating the wavefunctions of the $L$ band by the spin levels $|\alpha S_A S_B S_C S
M\rangle$, where $S_A$, $S_B$, and $S_C$ are the total spins of the three AFM sublattices, and $S_A$ = $S_B$
= $S_C$ = $S_i N/3$ ($\alpha$ denotes intermediate spin quantum
numbers).\cite{Bernu92,Bernu94,Schnack01,OW_SPINDYN} This is equivalent to describing the $L$ band by the
three-sublattice Hamiltonian $\hat{H}_{ABC} = -6J/N ( \hat{\textbf{S}}_A \cdot \hat{\textbf{S}}_B +
\hat{\textbf{S}}_B \cdot \hat{\textbf{S}}_C + \hat{\textbf{S}}_A \cdot \hat{\textbf{S}}_C)$.

The $E$ band, which corresponds to excitations out of the $L$ band, is then related to the spin levels with
$S_A$ = $S_B$ = $S_i N/3$, $S_C = S_i N/3$$-$$1$, and the permutations
thereof.\cite{Bernu94,OW_SPINDYN,Schnack01} These states are degenerate for $\hat{H}_{ABC}$, but are expected
to be split by the perturbation $\hat{H} - \hat{H}_{ABC}$. This suggests, as a first approximation, to
diagonalize $\hat{H}$ in the subspace of the spin levels $|\alpha S_A S_B S_C S M\rangle$, with $S_A$, $S_B$,
and $S_C$ chosen as appropriate for the $L$ and $E$ bands. In this work, where only the zero-temperature
excitation spectrum is considered, $\hat{H}$ was diagonalized in the subspace of the $S$=0 ground state and
the $S$=1 states of the $L$ and $E$ bands.\cite{Seq0} According to the coupling rules of three spins, the
$S$=0 ground state is non-degenerate, and the $S$=1 sectors of the $L$ and $E$ band consist of 3 and 81 spin
levels, respectively. The numerical calculation of the matrix elements $\langle \alpha S_A S_B S_C S
M|\hat{H}|\alpha S_A S_B S_C S M\rangle$ is simple thanks to the irreducible tensor operator
techniques.\cite{Mes85}

Sublattice Hamiltonians, such as $\hat{H}_{ABC}$, may be "derived" by Fourier analysis of the full
Hamiltonian.\cite{Anderson52,Bernu92,Bernu94} This procedure works for spin clusters with large symmetry,
such as rings, polyhedrons, or extended lattices. It may be constructed also by replacing each spin by the
mean-field spin $\hat{\textbf{S}}_{\eta} = 1/N_\eta \sum_{i \in \eta} \hat{\textbf{S}}_i$ of its
corresponding sublattice $\eta$, which provides a more general method.\cite{OW_Mn3x3NVT}

%%%%%%%%%%%%%%%%%%%%%%%%%%%%%%%%%%%%%%%%%%%%%%%%
%
% Results for Fe30 %

\begin{figure}
\includegraphics{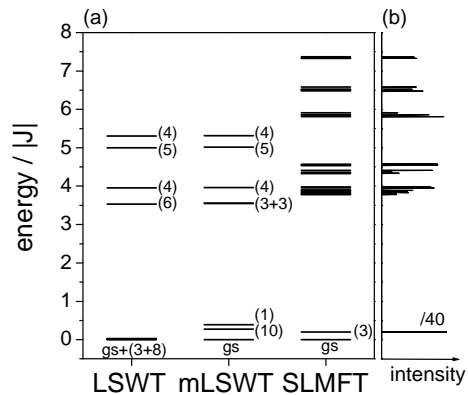}
\caption{\label{fig:1} (a) Energy levels of Fe$_{30}$ as calculated by LSWT, mLSWT, and SLMFT (gs = ground
state; numbers in brackets give the degeneracy). (b) $T$ = 0 neutron scattering intensity calculated by
SLMFT. The intensity of the peak at 0.2$|J|$ was divided by a factor of 40.}
\end{figure}

The results of the calculations for the excitations in Fe$_{30}$, as obtained by LSWT, mLSWT, and SLMFT, are
displayed in Fig.~\ref{fig:1}(a) (the mLSWT results were obtained previously in [\onlinecite{Cepas05}]). In
LSWT, the excitation spectrum consists of 3 levels at zero energy (with respect to the ground state),
followed by 8 levels in the energy range 0.0078 to 0.0247$|J|$, and four bands in the energy range
3.5$-$5.5$|J|$. As expected, mLSWT strongly affects the 11 lowest excitations, the spectrum is in particular
gapped. The first 10 excitations are at 0.2758$|J|$, followed by one at 0.3900$|J|$. The higher-lying
excitations are affected by less than 0.3\% (the levels at ca. 3.5$|J|$ are weakly split into two subgroups).
Clearly, the 11 low-lying excitations, and the ground-state, should be related to the $L$ band, while the
remaining higher-lying excitations should be related to the $E$ band. SLMFT produces a more structured energy
spectrum. It consists of the three triplets of the $L$ band at 0.2$|J|$, and the 81 triplets of the $E$ band,
which span the range 3.7811 to 7.3682$|J|$. For comparison, $\hat{H}_{ABC}$ gives energies of 0.2 and
5.2$|J|$ for the $L$ and $E$ band, respectively.

\begin{figure}
\includegraphics{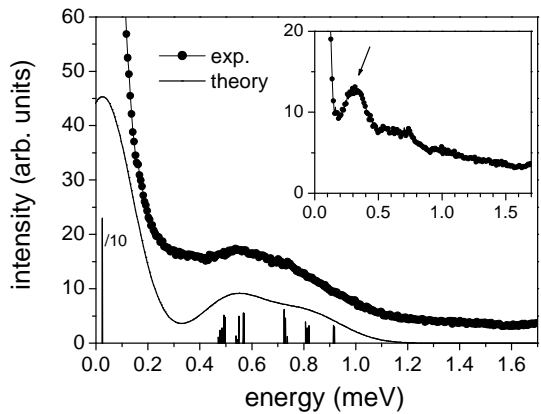}
\caption{\label{fig:2} Inelastic neutron scattering spectrum for Fe$_{30}$. The solid circles show the $T$ =
65~mK experimental data of [\onlinecite{Garlea06}] (data were recorded at OSIRIS with final neutron energy
$E_f$ =1.845~meV and integrated over the range $Q$ = 0.9$-$1.8~{\AA}$^{-1}$). The solid line and the vertical
lines represent the $T$ = 0 scattering intensity as calculated from SLMFT ($J$ = $-$0.125~meV; the solid line
was obtained by a convolution with a Gaussian of linewidth 0.27~meV). The inset displays the experimental
data after substraction of the theoretical curve. The arrow marks intensity not accounted for by the theory.}
\end{figure}

All three theories yield a significant splitting of the $E$ band, which is about 2$|J|$ for the SWTs and
3.5$|J|$ for the SLMFT. The experimental INS data at 65~mK, which is reproduced in Fig.~\ref{fig:2}, in fact
shows a broad feature from 0.2 to 1.1~meV. Within the three-sublattice Hamiltonian approach it was modeled by
a single Gaussian of width 0.66~meV.\cite{Garlea06} Its position was determined to 0.56~meV, which
corresponds to $J$ = $-$0.108~meV, in rough agreement with the value $J$ = $-$0.134~meV inferred from
magnetic susceptibility.\cite{Muller01,Garlea06} The above theories partially account for the observed
broadening (as noted before for the SWTs in [\onlinecite{Cepas05,Garlea06}]). Since SLMFT produces the
largest splitting, it is compared to experiment in more detail. The powder INS intensity integrated over $Q$
= 0.9$-$1.8~{\AA}$^{-1}$ was calculated from the wavefunctions obtained by SLMFT using the formulas of
[\onlinecite{OW_INS}] and the crystallographic Fe positions of Fe$_{30}$.\cite{Muller99} The result for $J$ =
$-$0.125~meV is shown as the vertical lines in Fig.~\ref{fig:2}. Deconvolution of this spectrum with a
Gaussian of width 0.27~meV produced the solid curve shown in Fig.~\ref{fig:2}. The theoretical spectrum
exhibits a strong peak at 0.025~meV due to the excitations from the ground state to the $L$-band triplets. It
is buried under the large quasielastic contribution and hence not detected in the experiment. Interestingly,
the experimental data in the range 0.4 to 1.1~meV is very well reproduced by SLMFT, and the obtained value
for $J$ agrees within 7\% with the value from magnetic susceptibility, which is a significant improvement
over the value obtained with $\hat{H}_{ABC}$. However, in the range 0.2 to 0.4~meV there is additional
intensity. It might be described by an "appropriate" background, but considering the resolution function of
OSIRIS it should be assigned to a feature not correctly reproduced by the SLMFT. In fact, subtracting the
theoretical curve from the experimental data (see inset of Fig.~\ref{fig:2}) strongly suggests additional
scattering from Fe$_{30}$ in this range.

%%%%%%%%%%%%%%%%%%%%%%%%%%%%%%%%%%%%%%%%%%%%%%%%
%
% FINAL DISCUSSION %

For the discussion, first the conceptional differences between the SWTs and SLMFT are noted. The SWTs start
from a symmetry-broken ground state, while SLMFT works with spin levels and is intrinsically spin rotational
invariant. Hence, the relation between the states obtained by SWTs on the one side and SLMFT (or exact
results) on the other side is not always obvious. For a bipartite system, this is not a big problem. SWTs
produce $N$ excitations which relate to $N-$1 spin excitations in the $S$=1 sector, i.e., one $L$-band and
$N-$2 $E$-band triplets (for a system with $S$=0 ground state). There is hence an obvious correspondence. The
$L$ band is associated to the (quantized) rotation of the N\'eel vector, and the $E$ band to the (discrete)
spin-wave excitations.\cite{Anderson52,Bernu92,OW_SPINDYN,OW_Cr8,SWcomment}

For a system with three (or more) sublattices the situation is less clear. The SWTs again produce $N$
excitations, while SLMFT concerns with 3 $L$- and 3($N-$3) $E$-band triplets (for a system with three
sublattices and a $S$=0 ground state). Clearly, since there are many more of them, not all of the $E$-band
triplets can be spin-wave excitations. This has an important implication. The (zero-temperature)
spin-correlation functions are governed by the transition matrix elements $\langle 0|\hat{S}_i^\nu| n\rangle$
($|0\rangle$ denotes the ground state, $|n\rangle$ the $n$th excited level). Within the
sublattice-Hamiltonian approximation therefore only the transitions from the ground state to the triplets of
the $L$ and $E$ bands have non-zero intensity, while all other transitions have zero
intensity.\cite{OW_SPINDYN} That is, only the triplets of the $L$ and $E$ bands are relevant. On the other
hand, there is no general restriction on the intensities of these transitions. That is, all triplets of the
$E$ band are relevant. This is underlined by Fig.~\ref{fig:1}(b), which reproduces the INS intensity as
obtained from SLMFT. Comparison with Fig.~\ref{fig:1}(a) shows that all parts of the energy spectrum
contribute. Since the number of $E$-band triplets considerably exceeds the number of magnons, one necessarily
has to conclude that (for three and more sublattices) SWTs are intrinsically not capable of describing the
spin dynamics correctly, or completely, respectively.

These points are in favor of SLMFT. However, a comparison of SWTs, SLMFT, and exact results for even-numbered
AF rings revealed that SLMFT provides an overall estimate for the spin-wave energies, but fails to reproduce
their dispersion (which SWTs of course do).\cite{OW_SWT_rings} On the other hand, the splitting of the $E$
band in AF finite chains due to the open boundary conditions is well grasped.\cite{Cr6hs} Apparently, SLMFT
accounts for some but not all features of the energy spectrum.

Concerning SLMFT, the studies on the rings and chains showed the following trends. The upper edge of the
exact $E$-band spectrum was obtained reasonably well; it was overestimated by about 10\%. The dispersion of
the magnons, however, was greatly underestimated, yielding very narrow $E$ bands: The exact $E$-band
spectrum extends to much lower energies as obtained by SLMFT. These trends could explain the discrepancies in
the SLMFT analysis of the Fe$_{30}$ INS data. First, a $J$ value which is 7\% too small is consistent with
the the trend to overestimate. Second, the scattering intensity in the range 0.2 to 0.4~meV could be
associated to magnons, which are at lower energies than calculated because SLMFT does not account for their
dispersion.

Concerning the SWTs on the other hand, they were found to reproduce the structure of the $E$ band very well
for the rings and chains, but to underestimate energies, i.e., to provide lower bounds for the $E$-band
spectrum (the accuracy of LSWT and mLSWT was ca. $-$20\%, that of interacting SWT and modified interacting
SWT about $-$10\%). With respect to Fe$_{30}$, Figure~\ref{fig:1}(a) then would imply that there are no
excitations below 3.5$|J|$ (besides the $L$-band excitations at very low energies), contradicting an
association of the 0.2$-$0.4~meV intensity to magnons. For the triangular AFM lattice, however, the inclusion
of 3rd- and 4th-order boson terms was found recently to result in a substantial suppression of the magnon
energies.\cite{Starykh06} Since a similar behavior is expected for Fe$_{30}$, the experimental intensity at
0.2$-$0.4~meV should be indeed associated to magnons, consistent with the above conclusion from SLMFT.

The calculated ground-state energies are $-$220.802$|J|$, $-$220.095$|J|$, and $-$195$|J|$ for the LSWT,
mLSWT, and SLMFT, respectively. A recent DMRG calculation yielded $-$211(2)$|J|$.\cite{Exler03} As observed
also for the rings and chains, LSWT yields a lower ground-state energy than mLSWT, while SLMFT is too high.
This is linked to the fact that SLMFT does not describe the spin-wave dispersion, and hence misses the
related quantum fluctuations. That LSWT and mLSWT, however, yields ground-state energies well below the DMRG
result should be considered another hint for the problems of the SWTs.

%%%%%%%%%%%%%%%%%%%%%%%%%%%%%%%%%%%%%%%%%%%%%%%%
%
% CONCLUSION %

In conclusion, three theories, LSWT, mLSWT, and SLMFT, have been tested for Fe$_{30}$. The SWTs, as they do
not cover all relevant excitations of the $E$ band, are not able to describe the spin dynamics, as measured
by INS for instance, correctly. SLMFT was successful to some extend. It reproduces a significant part of the
experimental INS spectrum, and yields a $J$ value which is in substantially better agreement with the $J$
value from magnetic susceptibility than that obtained with $\hat{H}_{ABC}$. However, it does not account for
all the experimentally observed excitation intensity.

For bipartite systems the combination of the results of SWTs and SLMFT yields a rather good picture of the
ground state and low-temperature excitations, even in a quantitative sense.\cite{OW_SWT_rings} For the
three-(and more) sublattice systems, these techniques could be considered useful to some extend for the
analysis of experimental data, in particular in the absence of better methods. Conceptionally, however, they
all are unsatisfactory. SWTs, on the one side, break the symmetry, which one has then to correct for somehow
afterwards. Furthermore, they do not reproduce the complete $E$ band. SLMFT, on the other hand, while
overcoming these two issues, is not able to yield the dispersion of the magnons correctly.

In order to further explore the situation for the three-(and more) sublattice systems, a comparison of the
SWTs and SLMFT with exact results would be crucial. The spin-1/2 analogue of Fe$_{30}$ can be handled by
exact diagonalization.\cite{Schmidt05} The experience for bipartite systems, however, indicates a breakdown
of the $L$/$E$-band concept for spin-1/2; hence the findings might not be transferable to systems with larger
spins.\cite{OW_SPINDYN} A comparison with the triangular AFM lattice (TL) is thus suggested, as the concepts
relevant here and for Fe$_{30}$ are basically identical.\cite{Bernu92,Bernu94}

Interestingly, the $N$=9 TL is exactly solved by $\hat{H}_{ABC}$, with proper values for $S_A$, $S_B$, and
$S_C$ (a similarity in sequence is noted: for bipartite systems $\hat{H}_{AB}$ is exact for the dimer,
square, and $N$=8 2D square lattice, for three-sublattice systems $\hat{H}_{ABC}$ is exact for the trimer,
octahedron, and $N$=9 TL). Hence, the $L$/$E$-band concept, and all its implications, is exact here. For
rings, which may be regarded as extensions of the dimer and square, the approximation by $\hat{H}_{AB}$
becomes the better the larger $S_i$ and the smaller $N$.\cite{OW_SPINDYN} For TLs, which may be regarded as
extensions of the trimer and $N$=9 TL, one may speculate about a similar trend of the accuracy of
$\hat{H}_{ABC}$ with $S_i$.

It would be particularly interesting to study the spin-5/2, $N$=12 TL, because exact diagonalization should
be possible (though not on the author's computers). The spin-1/2 case was deeply investigated
previously.\cite{Bernu92,Bernu94} Here it is found that the three $L$-band triplets are slightly split, and
that the splitting of the $L$ band increases with increasing $S$. This should be expected to happen also for
the spin-5/2 case, and for Fe$_{30}$. This effect could be relevant to explain the observed unusual field
dependence of the INS intensity:\cite{Garlea06} with increasing field the thermal population of the $L$-band
states would change, so that the relative contributions of the $E$-band excitations would change too.

%%%%%%%%%%%%%%%%%%%%%%%%%%%%%%%%%%%%%%%%%%%%%%%%
%
% APPENDIX %

%\appendix
%\section{}

%%%%%%%%%%%%%%%%%%%%%%%%%%%%%%%%%%%%%%%%%%%%%%%%
%
% ACKNOLEDGEMENTS %

\begin{acknowledgments}
The author thanks O. C\'epas and T. Ziman for the discussions on spin-wave theory, H.-J. Schmidt for the
suggestions concerning the calculation of classical ground states, and V. O. Garlea for providing the neutron
scattering data. Financial support by EC-RTN-QUEMOLNA (n$^\circ$ MRTN-CT-2003-504880) and the Swiss National
Science Foundation is acknowledged.
\end{acknowledgments}

%%%%%%%%%%%%%%%%%%%%%%%%%%%%%%%%%%%%%%%%%%%%%%%%
%
% REFERENCES %

%%%%%%%%%%%%%%%%%%%%%%%%%%%%%%%%%%%%%%%%%%%%%%%%
%
% END OF DOCUMENT %

\begin{references}
\bibitem{Seq0}
The $S$=0 states of the $E$ band start at an energy of 3.6735$|J|$. The INS intensity from the ground state
to these states is zero; they are hence not relevant at $T$=0. The temperature in the experiment suggests
some thermal population of the $S$=1 $L$-band states, which would produce INS intensity starting at
3.6735$|J|$$-$0.2$|J|$ = 0.43~meV. This mechanism hence cannot explain the observed INS intensity in the
range of 0.2$-$0.4~meV.
\bibitem{SWcomment}
There is a subtlety in the notation. The spectrum of standard SWTs exhibits zero-energy (Goldstone) modes,
which in the context of extended magnets are included in the notation of spin waves/magnons. With respect to
finite-size systems, however, notation should be changed. The zero modes (which then become the tower of
states=quasi-degenerate joint states=$L$ band)\cite{Anderson52,Bernu92,Bernu94,OW_SPINDYN,OW_Cr8} should not
be named spin waves, as they are of different physical significance. A notational separation of the
excitation spectrum into rotation of the N\'eel vector and the spin-waves/magnons (plus some higher-lying
excitations) is appropriate from several perspectives (the zero-modes cause the troubles in finite-size SWTs,
spin levels for the $L$ and $E$ bands have different sublattice lengths, scaling with $N$ is different,
etc.).
\end{references}
\end{document}